\documentclass{WileyMSP-template}
	
	\usepackage{graphicx}% Include figure files
	\usepackage{dcolumn}% Align table columns on decimal point
	\usepackage{bm}% bold math
	\usepackage{color}
	\usepackage{enumerate}
	\usepackage{braket}
	\usepackage{amsmath,amssymb,amsfonts}
	\usepackage{bbm}

	\usepackage[utf8]{inputenc}
	\usepackage[T1]{fontenc}
	\usepackage{mathptmx}
	
	\usepackage{textcomp} % without got an error on \textquotesingle

\begin{document}

	\pagestyle{fancy}
	%\rhead{\includegraphics[width=2.5cm]{vch-logo.png}}

	\title{Bright electrically controllable quantum-dot-molecule devices fabricated by in-situ electron-beam lithography}
	
	\maketitle
	
	\author{Johannes Schall}
	\author{Marielle Deconinck}
	\author{Nikolai Bart}
	\author{Matthias Florian}
	\author{Martin von Helversen}
	\author{Christian Dangel}
	\author{Ronny Schmidt}
	\author{Lucas Bremer}
	\author{Frederik Bopp} 
	\author{Isabell Hüllen}
	\author{Christopher Gies}
	\author{Dirk Reuter} 
	\author{Andreas D. Wieck}
	\author{Sven Rodt}
	\author{Jonathan J. Finley}
	\author{Frank Jahnke}
	\author{Arne Ludwig}
	\author{Stephan Reitzenstein}

	\begin{affiliations}
	J. Schall, M. Deconinck, M. von Helversen, R. Schmidt, L. Bremer, Dr. S. Rodt, Prof. S. Reitzenstein\\
	Institute of Solid State Physics\\
	Technische Universität Berlin\\
	Hardenbergstraße 36, D-10623 Berlin, Germany\\
	Email Address: stephan.reitzenstein@physik.tu-berlin.de
	
	N. Bart, Dr. A. Ludwig, Prof. A. D. Wieck\\
	Lehrstuhl für Angewandte Festkörperphysik\\
	Ruhr-Universität Bochum\\
	Universitätsstraße 150, D-44780 Bochum, Germany\\
	
	Dr. M. Florian, I. Hüllen, Dr. C. Gies, Prof. F. Jahnke\\
	Institute for Theoretical Physics\\
	University of Bremen\\
	P.O. Box 330 440, 28334 Bremen, Germany\\
	
	C. Dangel, F. Bopp, Prof. J. Finley\\
	Walter Schottky Institut and Physik Department\\
	Technische Universität München\\
	Am Coulombwall 4, 85748 Garching, Germany\\
	
	D. Reuter\\
	Department Physik\\
	Universität Paderborn\\
	Warburger Straße 100, 33098 Paderborn, Germany
	
	\end{affiliations}

% Keywords: Please provide a minimum of three and a maximum of seven keywords, separated by commas


\justifying

\begin{abstract}
\noindent Self-organized semiconductor quantum dots represent almost ideal two-level systems, which have strong potential to applications in photonic quantum technologies. For instance, they can act as emitters in close-to-ideal quantum light sources. Coupled quantum dot systems with significantly increased functionality are potentially of even stronger interest since they can be used to host ultra-stable singlet-triplet spin qubits for efficient spin-photon interfaces and for deterministic photonic 2D cluster-state generation. We realize an advanced quantum dot molecule (QDM) device and demonstrate excellent optical properties. The device includes electrically controllable QDMs based on stacked quantum dots in a pin-diode structure. The QDMs are deterministically integrated into a photonic structure with a circular Bragg grating using in-situ electron beam lithography. We measure a photon extraction efficiency of up to (24$\pm$4)\% in good agreement with numerical simulations. The coupling character of the QDMs is clearly demonstrated by bias voltage dependent spectroscopy that also controls the orbital couplings of the QDMs and their charge state in quantitative agreement with theory. The QDM devices show excellent single-photon emission properties with a multi-photon suppression of  $g^{(2)}(0) = (3.9 \pm 0.5) \cdot 10^{-3}$. These metrics make the developed QDM devices attractive building blocks for use in future photonic quantum networks using advanced nanophotonic hardware.
\end{abstract}

% Text: Please use section headings and subheadings as specified below. For communications, all section headings apart from Experimental Section should be removed
% Please make the first reference to a display item bold: \textbf{Figure 1}
% Do not abbreviate Figure, Equation, etc.; display items are always singular, i.e., Figure 1 and 2.
% Equations are always singular, i.e., Equation 1 and 2, and should be inserted using the {equation} environment, not as graphics
% Please do not use footnotes in the text, additional information can be added to the Reference list.

\section{Introduction}

In the field of photonic quantum technology, individual photons play a prominent role. As flying qubits, they serve primarily as information carriers for low-loss quantum communication over long distances~\cite{Briegel1998, Lo_1999, Hughes_2000, Duan_2001, Korzh2015, Yin2020}. The information to be transmitted is typically encoded into the polarization of the photons~\cite{BB84, Jennewein_2000}. In the case of quantum repeater networks, but also for future distributed quantum computers and global quantum networks, it is of central importance to temporarily store and retrieve the quantum information to be transmitted for as long as possible in the form of stationary qubits in quantum memories \cite{Lvovsky_2009, Afzelius_2015}. 
In this context it is a great challenge to develop device concepts that simultaneously have a high level of performance in terms of single-photon generation with high rate and high multiphoton suppression and that are also suitable as efficient quantum memories. The NV-center in diamond, for example, has a very long spin coherence time, which makes it ideal as a quantum memory. However, these centers show high Huang-Rhys coupling factors, such that only about 3\% of radiative emission occurs via the preferred zero-phonon line transition. This is highly problematic with respect to on-demand single-photon generation with high photon flux~\cite{Jelezko_2006}. \\

In contrast, self-assembled InGaAs and GaAs quantum dots (QDs) are excellent single-photon emitters with almost negligible multi-photon emission probability \cite{Schweickert2018}, close to ideal indistinguishability \cite{Wei2014} and photon extraction efficiencies exceeding 80\% \cite{Liu_2019}. 
However, these nanostructures have relatively short spin coherence times, which has a deleterious effect on their possible use as quantum memories when relying on the storage of single carriers \cite{Kroutvar_2004, Young_2007, Heiss_2008}. It has recently been shown that the spin coherence times can be increased to about 2 $\mu$s by all-optical Hahn echo decoupling \cite{Stockill_2016}. However, such approaches may complicate protocols due to the infidelity of control pulses. To circumvent this problem, one can go one step further and not work with individual QDs, but rather look at potentially more powerful concepts that are based on singlet-triplet qubits in quantum dot molecules (QDMs) - This approach promises storage times in excess of 1 ms~\cite{Boyerdela2011}. In this concept the electric field dependent charge separation in a QDM is used to initialize and store an exciton-spin state, before the readout via fast radiative recombination is triggered by a suitable external voltage pulse. In addition, coupled QDs are also very interesting nanostructures for the generation of two-dimensional cluster states of polarization encoded photonic qubits \cite{Economou_2010}.\\

InGaAs QDMs with high optical quality were first implemented in 2001 and optically analyzed with regard to their coupling behavior \cite{Bayer_2001}. 
The thickness of the tunnel barrier, which separates the lower and upper QDs, was varied during growth and the resulting energy splitting between binding and anti-binding hybrid states was statistically investigated. Building on this, QDMs have been studied extensively with regard to their coupling and charge carrier storage properties \cite{Shtrichman_2002, Gerardot_2005, Scheibner_2007, Ardelt_2016,  Jennings_2019}. Here, the external control of the electronic properties via an electrical field in doped and contacted structures is essential for these QDM implementations. For example, the important coherent tunnel-coupling between resonant electronic states of the lower and upper QDs to form hybridized molecular orbits has been clearly demonstrated via anti-crossings observed in electric-field dependent optical spectra~\cite{stinaff_optical_2006}.\\

An important aspect in this context, and with regard to possible applications of QDMs in photonic quantum technology, is the efficient coupling between stationary and flying qubits. For quantum photonic applications, in addition to precise control of the electronic states in QDMs, highly efficient spin-photon coupling must also be guaranteed to be able to map spin qubits in a QDM to the polarization of emitted photons and to transmit it to a quantum network \cite{Afzelius_2015}. For the purpose of photonic coupling, a QDM was recently integrated into a photonic crystal nanocavity \cite{Vora_2019}. Due to the high quality (Q) factor and the low mode volume of the cavity, a strongly coupled QDM-cavity system in the cavity quantum electrodynamics (cQED) regime could be demonstrated. Although this coupling is of interest for the implementation of spin-photon interfaces, the narrow-band high-Q character of the cavity mode prevents simultaneous coupling to the interband transition in both QDs forming the molecule. \\

In recent years, broadband approaches to increase photon extraction efficiency have been established, which include photonic wires~\cite{Claudon_2010}, microlenses~\cite{Gschrey_2015} and circular Bragg gratings (CBGs)~\cite{Davan_o_2011, Liu_2019}. 
In the case of the very attractive CBG concept, they almost ideally combine ultra-high photon extraction efficiency with a moderately high and easily tunable light-matter interaction~\cite{Liu_2019}. So far, approaches of this kind have only been used for individual QDs, which is partly due to the fact that they are difficult to reconcile with electrical field control, which is inevitably required for QDM quantum photonic devices. Until now, enhanced photon extraction for electrically tunable QDMs has only be reported in lateral QDMs in a DBR cavity without deterministic device processing~\cite{hermannstadter_2009}.\\

In this work we report on the development of electrically tunable single-QDM devices with strongly enhanced photon extraction efficiency, the potential of selective optical charging~\cite{Kroutvar_2004, Heiss_2008} and electrical control of spin-spin interaction in the ultra coherent singlet-triplet basis. The devices are based on vertically stacked InGaAs QDMs embedded in pin-diode structures that were grown using molecular beam epitaxy (MBE). Suitable QDMs were selected using in-situ electron beam lithography (EBL) and deterministically integrated into photonic structures. In the underlying device design, the QDMs are electrically controlled via n- and p-contacts near the surface in a planar design. An increased photon extraction efficiency is achieved by combining a back-side distributed Bragg reflector (DBR) with an upper ring resonator, the design of which has been optimized using the finite element method (FEM). 
The functionality of the QDM devices is studied spectroscopically in order to demonstrate not only the increased photon extraction efficiency, but also the coupling character of the QDMs and their single photon emission.\\

\section{Results}

\subsection{Device Design}

\begin{figure}
\begin{center}
\includegraphics[width=0.55\linewidth]{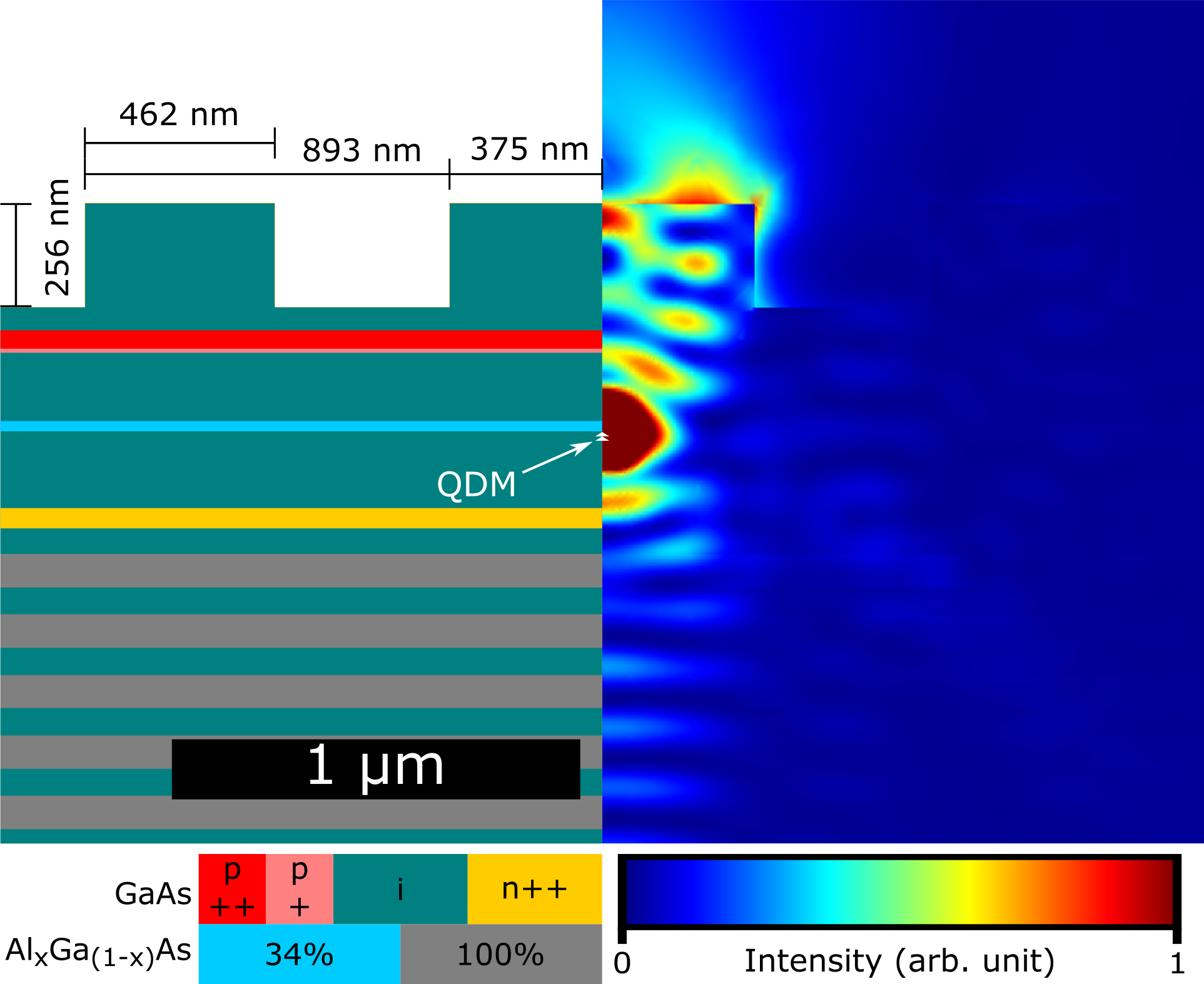}  
\caption{Schematic representation of the device design (left) together with the simulated electric field distribution (right). The device contains a back-side DBR with 23 $\lambda/4$-thick mirror pairs and QDMs, which are embedded in a pin-diode structure with highly doped n$^{++} $ and p$^{++}$ layers for electrical contacting. A CBG consisting of a central mesa with four rings (see also Fig.~\ref{fig:Fig_2}(c)) is structured on the surface in order to maximize the photon extraction efficiency $\eta_{ext}$ for a collecting optics with NA = 0.8 (only one ring is shown here for clarity). The numerical FEM simulation yields $\eta_{ext}$ = 24.4 \% for the specified layer thicknesses and material compositions. More details on the sample structure are given in the main text.}
\label{fig:Fig_1}
\end{center}
\end{figure}

The overarching goal of this work is to develop and manufacture electrically operated QDM devices with high coupling efficiency for future quantum memory applications in photonic quantum technology. For this purpose, we design, fabricate and study a QDM device with intracavity contacts and a circular Bragg grating on top. The device design is optimized numerically to ensure a good balance between precise electrical field control of the QDM and high photon extraction efficiency. As illustrated in Fig.~\ref{fig:Fig_1}, our design envisages a near-surface GaAs pin-diode structure in the center of which the self-organized InGaAs QDMs are integrated. The device contains a p-side Al$_{0.34}$Ga$_{0.66}$As barrier, that serves to suppress hole tunneling. This design was chosen in order to ensure a comparatively simple electrical contact layout that does not require any nanostructured top contacts that are complex to fabricate and potentially enhance the optical losses. This approach allows a flexible choice of the photonic design for enhanced photon extraction without having to pay attention to impeding contact structures. 
However, certain compromises have to be made with regard to the maximum achievable photon extraction efficiency, for example in comparison to CBG based single photon sources that are optimized solely for their optical properties and without the need for an electrically contacted diode structure. Classically, the quantum emitter is placed in the central mesa of the CBG structure. We decided to place the diode with the enclosed QDM below the CBG for enabling an easy and robust contacting scheme and the compatibility with deterministic device processing. 
In the numerical FEM simulations performed using JCMSuite \cite{Pomplun_2007, Schneider_2018}, we considered a backside AlAs/GaAs DBR and a surface CBG with the pin-doped central region and a QDM in between. 
Varying the layer design and the geometry of the CBG, we maximized the photon extraction efficiency for the experimental collection optics with an NA of 0.8, while maintaining suitable fabrication tolerances. The corresponding electric field distribution is presented on the right hand side of Fig.~\ref{fig:Fig_1}. The standard fabrication tolerances of our in-situ EBL system \cite{Gschrey_2015_2} like a lateral offset of the CBG and the QDM or a variance in the structural width can result in a reduction of the optimized extraction efficiency which accounts to 24.4\%.
For example, a lateral offset of 35 nm results in a decrease by 1.5\% and a deviation of the width of the central mesa and the rings of 50 nm decreases the efficiency by 6.7\%. The broadband efficiency of our structures enables an extraction efficiency better than 20\% within a 30 nm spectral range.
Our simulations show that $\eta_{ext}$ increases marginally when adding more than 3 rings. 
Therefore, the number of rings in the technological implementation was limited to four.\\

\subsection{Deterministic Device Fabrication}

Following the design specifications, the planar semiconductor heterostructure was grown using MBE on an undoped (100) oriented GaAs wafer as described in Sec.~\ref{growth}. A sample piece of 3 mm $\times$ 10 mm with a QD density of $(0.5 - 5) \cdot 10^{6}$ cm$^{-2}$ was selected. Its n- and p-doped layers were accessed near two corners of the sample by UV lithography and reactive ion etching for electrical contacting. For the n-contact, 20 nm Ni, 100 nm Au$_{0.88}$Ge$_{0.12}$ and 250 nm Au were deposited. The p-contact consists of layers of 20 nm Ti, 50 nm Pt and 250 nm Au. After contacting, a trench around the desired pin-diode area was processed by vertical etching and stopping right below the p-layer to isolate the diode from possible electrical short circuits at the edge of the sample. Finally, the diode structure was mounted onto a chip carrier and electrical connections were realized by wire bonding. \\

\begin{figure}
\begin{center}
\includegraphics[width=0.55\linewidth]{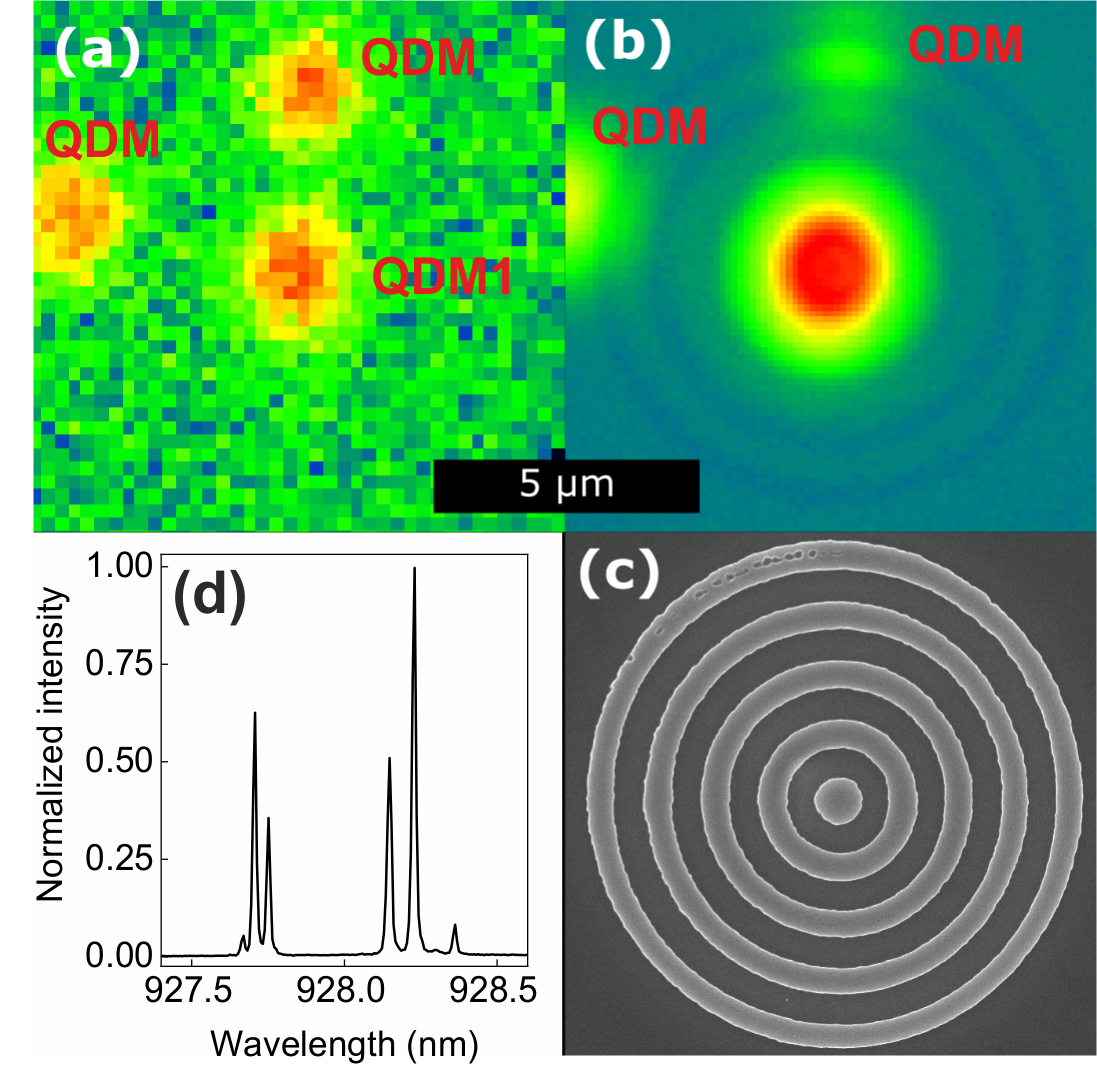}
\caption{(a) CL intensity map recorded during the in-situ EBL step. The CL map in (linear color scale) shows three bright QDM emission centers, of which QDM1 in the image center was selected for device integration, based on its brightness and the emission wavelength with respect to the numerically optimized CBG structure.  (b) CL intensity map of the same sample position after the deterministic integration of QDM1 into a CBG structure. Due to the photonic engineering and the CBG effect, the detected CL emission intensity of QDM1 is significantly higher than that of the other two QDMs. (c) Scanning electron microscope image of the QDM-CBG device with four concentric rings and a central mesa fabricated by in-situ EBL (same scale as (a) and (b)). (d) $\mu$PL spectrum of the manufactured QDM-CBG device.}
\label{fig:Fig_2}
\end{center}
\end{figure}

For the identification of promising QDMs and their on-spot integration below CBG structures we used the in-situ EBL nanotechnology concept. We used this nanophotonic technology platform previously for the fabrication of bright microlenses~\cite{Gschrey_2015}, mesas~\cite{Srocka_2020} and waveguide structures \cite{Schnauber_2018} using proximity-corrected gray scale patterns~\cite{Schnauber_2019} with deterministically integrated QDs. 
It should be noted here that functional QDMs cannot be identified definitely without performing bias-voltage dependent optical spectroscopy to probe for tunnel-coupling-related luminescence features. Such a time-consuming investigation and the related high electron dose per site is not compatible with the in-situ EBL process since the electron-beam sensitive resist would become overexposed during the spectral and bias-dependent cathodoluminescence (CL) mapping. Consequently, in the in-situ EBL process we concentrated on the selection of spatially isolated emission sites with high CL intensities (cf.~Fig.~\ref{fig:Fig_2}(a) with three QDM related emission spots). This together with the high QDM formation probability of $\approx$ 80 \% (see discussion of Fig.~\ref{fig:Fig_4}) leads to a high yield of functional QDM devices. The envisaged CBG structures were numerically optimized for an emission wavelength of 930~nm (see Fig.~\ref{fig:Fig_3}(b)). However, the broadband enhancement of our structures enables also the boost of extraction efficiency in the wavelength range around 925~nm at the center of the inhomogeneously broadened emission band of our QDMs. This way, stacked QDs (with high probability for acting as QDMs) were deterministically integrated into CBG structures as exemplarily shown in Fig.~\ref{fig:Fig_2}(b). This figure shows a CL map of the same sample area as presented in Fig.~\ref{fig:Fig_2}(a) after device processing. The deterministic integrated QDM1 shows significantly higher CL intensity than the other two QDMs due to the CBG-enhanced photon extraction.\\

\subsection{Electric Field Controlled Tunnel Coupling of Quantum Dot Molecules}

\begin{figure}
\begin{center}
\includegraphics[width=0.55\linewidth]{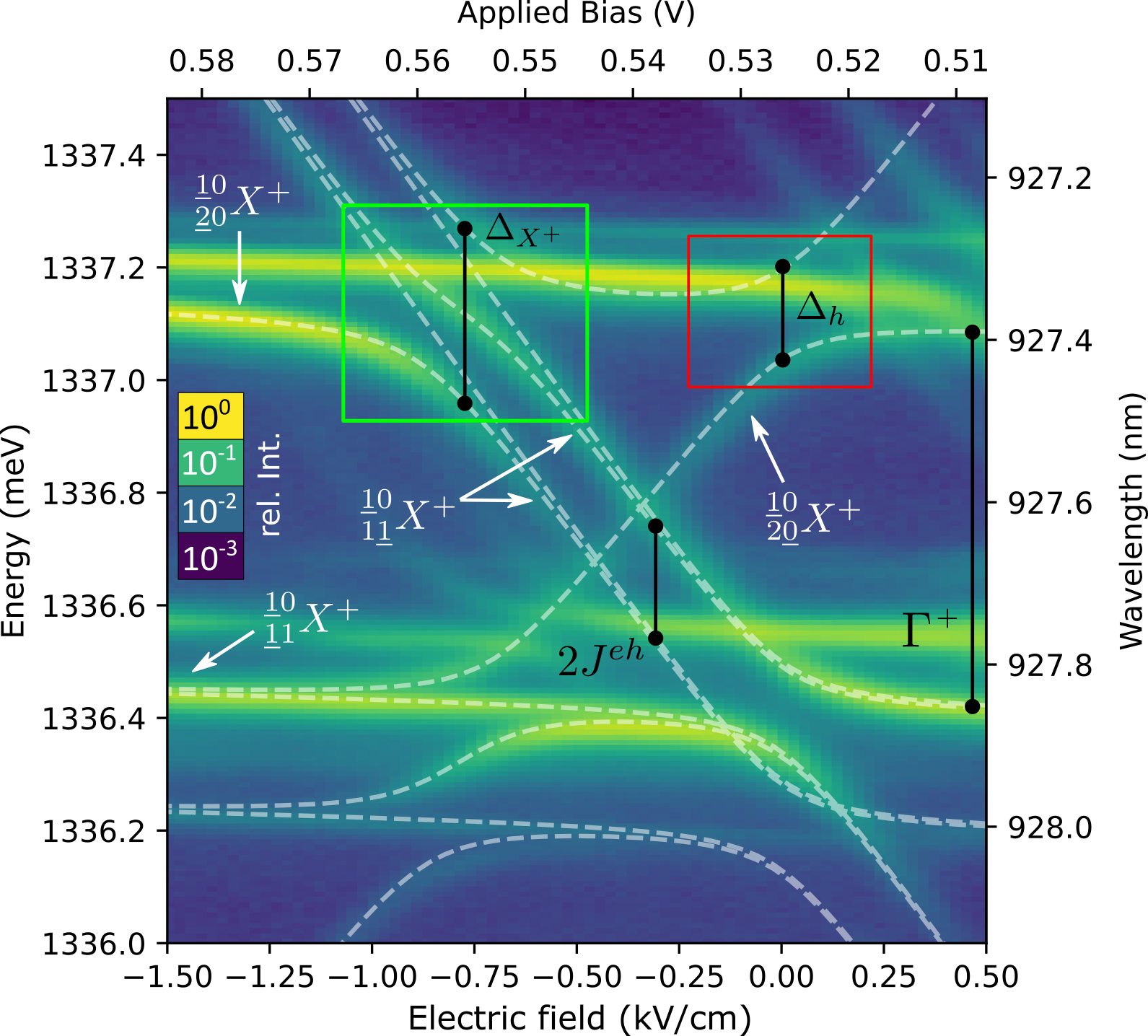}
\caption{Bias voltage dependent $\mu$PL intensity map showing the spectral tuning and coupling of excitonic states of QDM1. Tunnel coupling and the existence of hybridized binding and anti-binding states of the QDM are identified by the anti-crossing behavior (indicated by the green and red rectangles). The associated coherent mode-splitting of $\Delta_h = 2t_h = 160~\mu eV$, is consistent with the $7.3$~nm thick tunnel barrier. The experimental data is described with excellent quantitative agreement by our theory, which allows us to assign the observed emission features to charge states and radiative transitions of the QDM as indicated in the graph. Moreover, it provides important information about the QDM coupling parameters ($2J^{eh} = 210~\mu eV$, $\Gamma^{+} = \Gamma - J^{eh} = 680~\mu eV$, $\Delta_{X^+}=2\sqrt{2t_h^2 + (J^{eh})^2} = 310~\mu eV$). The field strength is plotted relative to the point at which the hole ground state resonance occurs. Trion states are labeled $^{e_Be_T}_{h_Bh_T}X^+$, where the left superscripts (subscripts) denote the number of electrons (holes) in the bottom and top QD. Optical transitions are labeled by underlining the recombining carriers. Following the notation of Ref.~\cite{stinaff_optical_2006} the QDM states are labeled $\mathrm{^{e_B,e_T}_{h_B,h_T}X^Q}$, where the left superscripts (subscripts) give the number of electrons (holes) in the bottom $\mathrm{e_B}$ ($\mathrm{h_B}$) and top $\mathrm{e_T}$ ($\mathrm{h_T}$) dots and Q is the total charge of the system. The transitions are indicated by underlining the recombining carriers.}
\label{fig:Fig_4}
\end{center}
\end{figure}

After deterministic device fabrication and post-characterization performed using CL mapping, the QDM devices were further examined via optical spectroscopy. First, $\mu$PL measurements were carried out under non-resonant excitation with a 80 MHz pulsed 860 nm laser at a temperature of 4.2 K. A corresponding $\mu$PL spectrum of the device (CBG with integrated QDM1) previously examined using CL is shown in Fig.~\ref{fig:Fig_2} for zero bias voltage. The spectrum shows clear single QD features with narrow, resolution-limited linewidths of $\approx$ 12 pm (20~$\mu$eV). The observed emission lines can be assigned to direct and indirect excitons of the QDM as detailed below when discussing Fig.~\ref{fig:Fig_4}.\\

For the intended applications in photonic quantum technology, it is crucial to verify the coherent coupling of the electronic states in stacked QDs and to control it externally. For this purpose, a voltage applied to the external contacts of the pin-diode generates an electrical field in the device along the growth direction. In this configuration, a bias-voltage dependent $\mu$PL map shows optical transitions with large electric field dependencies (Stark shifts) as well as a crossing and anticrossing pattern that can be clearly seen in Fig.~\ref{fig:Fig_4} for QMD1.
Emission of a direct exciton ($^{\underline 1,0}_{\underline 1, 0}X^0$), which arises from an electron and hole confined primarily in the same QD, shows a significantly weaker shift compared to the indirect exciton ($^{\underline 1,0}_{0,\underline 1}X^0$) (please see Fig.~\ref{fig:Fig_4} and Sec.~\ref{theory} for the nomenclature and details of the underlying model). Here, electron and hole states are located in different QDs~\cite{stinaff_optical_2006} and the recombination obeys a strong linear Stark shift $\Delta E=e[d + (h_B + h_T)/2]F$ that depends on the barrier thickness ($d = 7.3$ nm). $F$ is the strength of the electric field and $h_B$ ($h_T$) the heights of the bottom (top) QD.
Anticrossings are observed when direct and indirect transition energies approach each other and a hole (electron) becomes delocalized across both QDs, forming bonding and antibonding molecular states. Here the width of the anticrossing splitting depends on the strength of QDM tunnel coupling. In addition to these basic signatures, Fig.~\ref{fig:Fig_4} shows a significantly more complex fingerprint of the coupled QDs with an intricate X-shape pattern with several anticrossing splittings. This property is due to the existence of charged excitonic states where a strong indirect transition ($^{\underline 1,0}_{1,\underline 1}X^+$) anticrosses two direct transitions ($^{\underline 1,0}_{\underline 2, 0}X^+$, $^{\underline 1,0}_{\underline 1, 1}X^+$)~\cite{stinaff_optical_2006}. In our layer design with a p-side tunnel barrier, positively charged excitonic states are preferably expected. Characteristic for singly charged exciton transitions is the singlet-triplet mixing with an apparent triplet transition that wiggles through the $X^+$ resonance (green box in Fig.~\ref{fig:Fig_4}) as well as the single hole anticrossing at the hole ground state resonance (red box in Fig.~\ref{fig:Fig_4})~\cite{Scheibner_2007}.

Overall, these spectral properties of the vertically coupled QDs clearly show the existence of QDM states, which, like the charge state of the coupled system, can be controlled by the externally applied bias voltage. It is worth noting that a statistical analysis of the emission properties shows that $\approx$ 80 \% of the devices contain stacked QDs with a suitable electronic structure to form QDMs via tunnel coupling at electric fields that are small enough to perform optical experiments. This makes the developed quantum devices very attractive candidates for spin-photon interfaces and 2D cluster-state generators.

\subsection{Photon-Extraction Efficiency of Quantum-Dot-Molecule Devices}

A central objective of this work is to realize QDM devices with high photon extraction efficiency. In order to evaluate the photon extraction efficiency of our approach 6 QDM devices with integrated CBGs (named QDM1-6) and 19 QDMs in an unstructured sample area as reference, were investigated by $\mu$PL measurements under non-resonant (860 nm) pulsed excitation with a laser repetition rate of 80 MHz. For each QDM the measurement was carried out at saturation pump power of the brightest excitonic line in a calibrated $\mu$PL setup with a detection efficiency of (4.6 $\pm$ 0.5)\%. In the case of QDM1, adding up the photon count rates of the relevant single-excitonic lines yields a photon-extraction efficiency of $(19.9\pm 2.4)$\% for a count rate of 730 kHz. \\

\begin{figure}
\begin{center}
\includegraphics[width=0.55\linewidth]{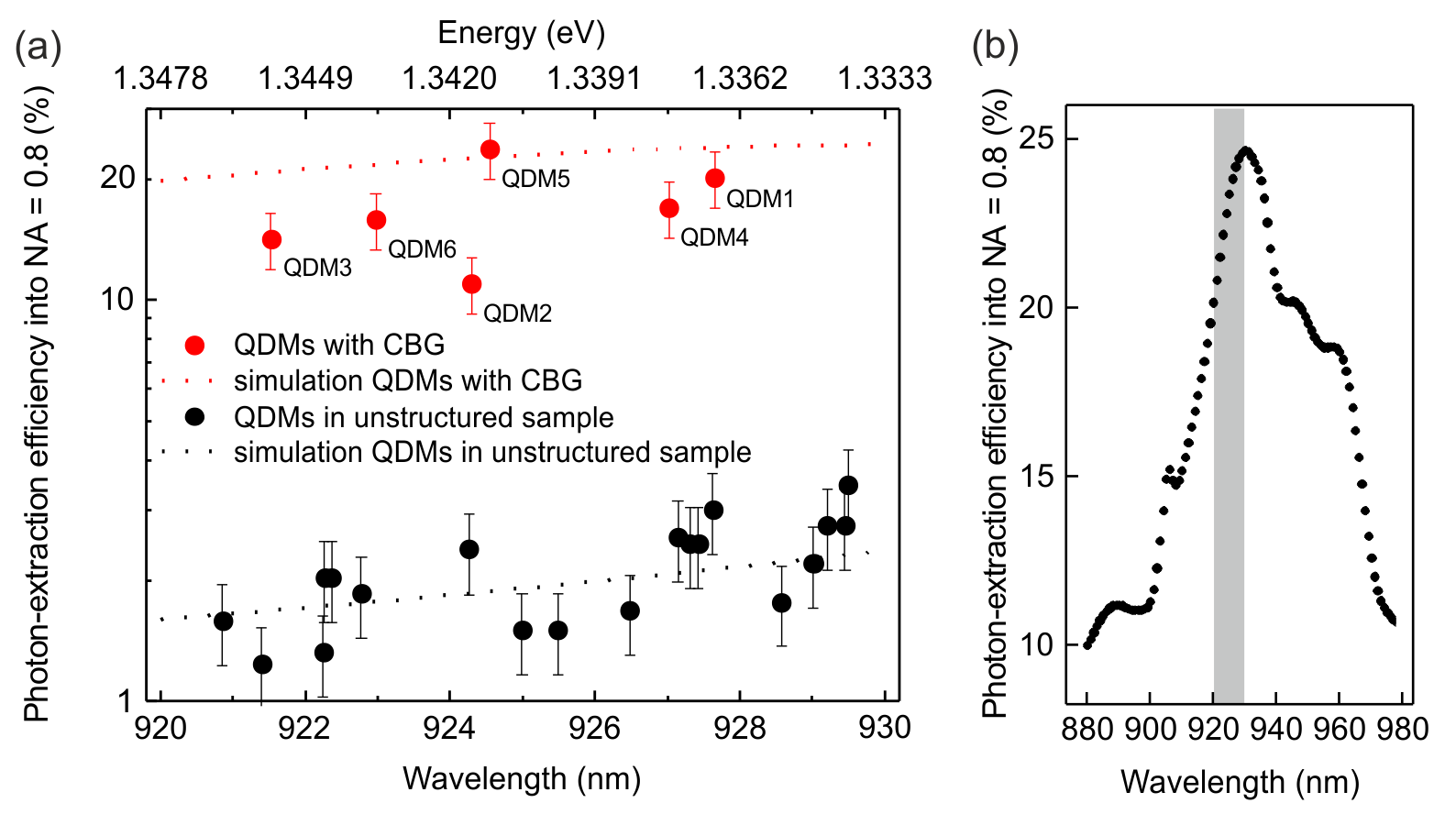}
\caption{(a) Photon-extraction efficiency of 6 QDM devices (QDM1-QDM6) with CBG and 19 reference QDMs in planar sample areas for emission wavelengths between 920 nm and 930 nm. The experimental data is in good agreement with the theoretical prediction (dashed lines) based on FEM simulations. The highest photon extraction efficiency of (24$\pm$4)\% was observed for QDM5. (b) Calculated photon-extraction efficiency (into NA = 0.8) of QDM devices in a wider wavelength range. The wavelength range corresponding to panel (a) is indicated by the grey area.}
\label{fig:Fig_3}
\end{center}
\end{figure}

The  determined photon-extraction efficiencies of all 6 devices (QDM1-6) are presented in Fig.~\ref{fig:Fig_3}(a). All devices show photon-extraction efficiencies larger than 10\% and a maximum value of (24$\pm$4)\% determined for QDM5 which nicely confirmed the high quality and yield of our deterministic fabrication process. It is interesting to compare the photon-extraction efficiencies of the CBG-QDM devices with those of the reference QDMs in the unstructured sample area. Corresponding emission rates and photon-extraction efficiencies of  these 19 unpatterned QDMs are also plotted in Fig.~\ref{fig:Fig_3}(a). They show that an enhancement of more than an order of magnitude is obtained by device integration into CBG structures. Additionally, we compare the experimental data with photon-extraction efficiencies obtained by FEM simulations with and without CBG (dashed and dotted lines). The efficiency values of QDMs in the unstructured sample area are in very good agreement with the trend of calculated values with a slight increase of photon-extraction efficiencies with increasing wavelength. The values of the CBG-QDM devices are in 5 cases below the expected maximum and match the expected value in the case of QDM5. These results show the overall very good process control, the good reproducibility of the technology and the reliability of the numerical simulations, whereby deviations from the expected photon-extraction efficiency to lower values in case of the CBG-QDMs are attributed to a slight lateral offset of the QDMs relative to the CBG and to structural imperfections of the CBGs. Fig.~\ref{fig:Fig_3}(b) presents the calculated photon-extraction efficiencies in a wider range of wavelengths. The data yields that the broadband efficiency of our structures enables an extraction efficiency better than approximately 20\% within a 30 nm spectral range and that the realized QDM devices emit at wavelengths (indicated by the grey area) close to the maximum of the efficiency curve. 

\subsection{Single-Photon Emission of Quantum Dot Molecules}

\begin{figure}[]
\begin{center}
\includegraphics[width=0.55\linewidth]{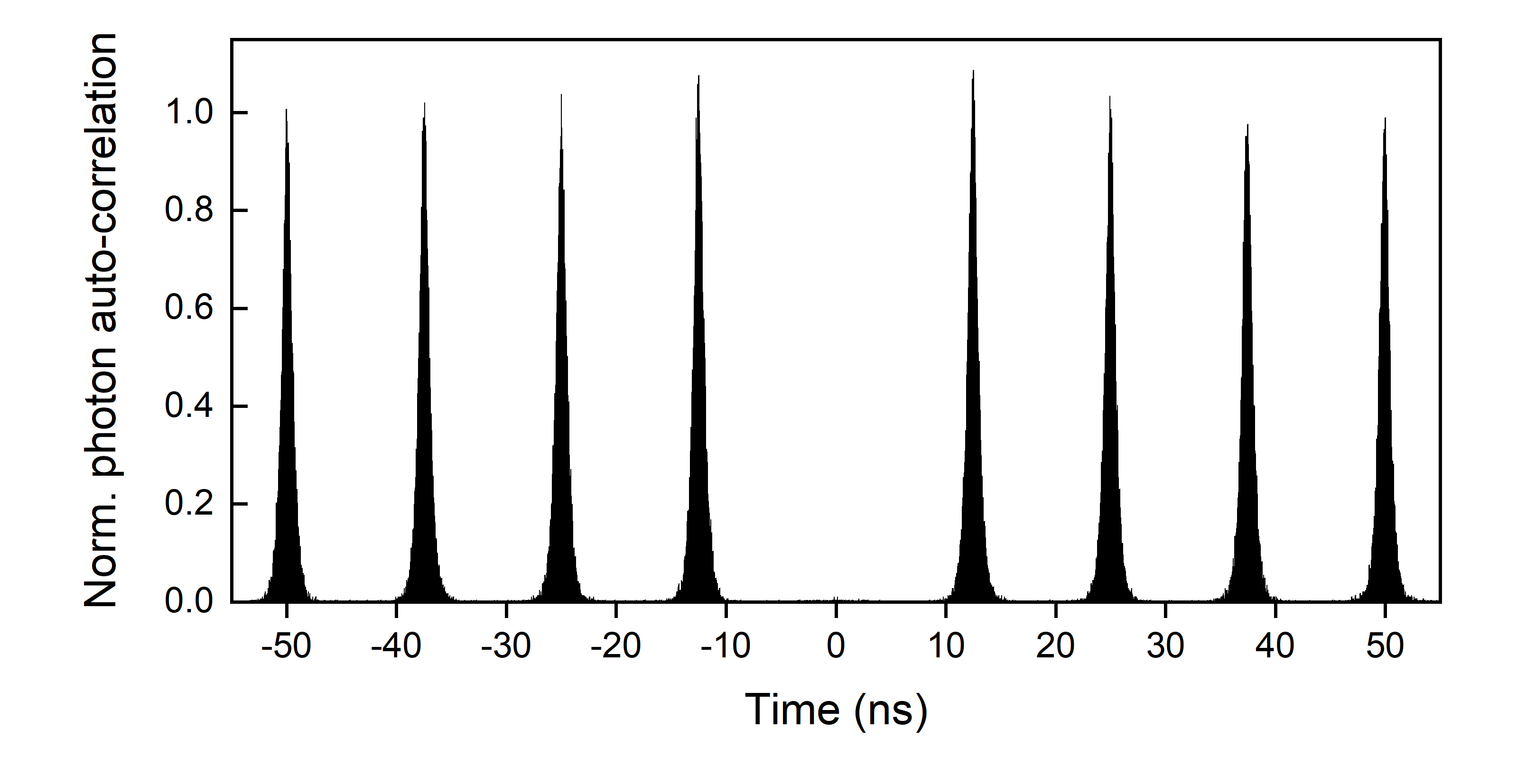}
\caption{Photon autocorrelation function of QDM1 under pulsed p-shell excitation. The investigated direct excitonic line shows close to ideal single-photon emission associated with $g^{(2)}(0) = (3.9 \pm 0.5) \cdot 10^{-3}$.}
\label{fig:Fig_5}
\end{center}
\end{figure}

In addition to the QDM coupling behavior and the high photon extraction efficiency, non-classical light emission is an important property envisioned for application of the developed devices in photonic quantum technology. In particular, it is important to demonstrate the single-photon emission character through photon auto-correlation studies. Corresponding measurements were carried out with a Hanbury Brown and Twiss setup equipped with superconducting nanowire single-photon detectors (SNSDPs) under pulsed p-shell excitation at 909.8 nm with a repetition rate of 80 MHz (T = 4.5 K). 
To study the photon statistics of the QDM-device emission we investigated a direct exciton transition of QDM1. We applied about 80\% of the saturation intensity and set a bias voltage of -0.1 V to study the emission line outside the intersection area. The corresponding normalized auto-correlation function is shown in Fig.~\ref{fig:Fig_5}. It clearly shows single-photon emission with a very high multi-photon suppression associated $g^{(2)}(0) = (3.9 \pm 0.5) \cdot 10^{-3}$. It is worth noting that this value was obtained directly by integrating the $g^{(2)} (\tau)$ function in the range -6.25 ns < $\tau$ < 6.25 ns of the central peak without any background subtraction. On the same QDM we performed auto-correlation measurements for a single line at 0.36 V and on the direct $\mathrm{^{\underline{1},0}_{\underline{1},1}X^{+}}$ transition at 0.8 V. Both measurements reveal very high multi-photon suppression associated with $g^{(2)}(0) = (5.9 \pm 0.5) \cdot 10^{-3}$ for 0.36 V and $g^{(2)}(0) = (4.0 \pm 4.0) \cdot 10^{-3}$ for 0.8 V, respectively.

For the other 5 QDM-CBGs, we conducted HBT measurements under wetting layer excitation and determined $g^{(2)}(0)$ values in the range of $(7.0 \pm 0.7) \cdot 10^{-3}$. We attribute the slightly higher values to the changed excitation condition (wetting layer instead of p-shell excitation). The $g^{(2)}(0)$ value of $(8.0 \pm 1.5) \cdot 10^{-3}$ for an QDM without a CBG structure shows, that the structuring does not influence the single-photon emission. The very high multi-photon suppression in all cases reflects again the high optical quality of the developed QDM devices, which makes them very attractive for applications in photonic quantum technology. 

\section{Conclusion}

In summary, we have developed QDM devices with excellent optical and quantum optical emission properties. The devices are based on vertically stacked self-assembled InGaAs quantum dots in a pin-diode heterostructure. Strong enhancement of the photon extraction efficiency with a maximum value of (24$\pm$4)\% is realized by a numerically optimized combination of a lower DBR mirror and an upper circular Bragg grating. This facilitates a detailed spectral observation of charge states and the coupling behavior of deterministically integrated QDMs precisely controlled by intra-cavity contacts as well as their description with excellent quantitative agreement by a microscopic theory. From a fundamental point of view, the combination of electrical-tunable and high-quality QDMs in deterministic CBG structures will enable more detailed spectroscopic investigations of tunnel-coupling-related phenomena in future works.  Beyond that, very high multi-photon suppression associated with $g^{(2)}(0) = (3.9 \pm 0.5) \cdot 10^{-3}$ makes the QDM devices very attractive building blocks for applications in photonic quantum technology such as quantum repeater based quantum networks and photonic cluster state generators.

\section{Methods}
\subsection{Sample Growth}
\label{growth}
The sample growth using molecular beam epitaxy starts with initial smoothing layers grown at 620°C consisting of 20 thin (2.5 nm) and annealed GaAs layers and a short period superlattice (19 $\times$ 2 nm AlAs and 2 nm GaAs). Next, a DBR was grown at the same temperature, consisting of 23 pairs of 67 nm GaAs and 81 nm AlAs. After a temperature reduction to 600°C, a 50 nm Si-doped GaAs layer with a dopant concentration of $2 \cdot 10^{18}\, / \, \textrm{cm}^{3}$ was grown, followed by 5 nm GaAs grown at 575°C to prevent Si segregation to subsequent layers. After a 163.6 nm GaAs buffer layer, InAs was deposited at 525°C in 11 cycles of 4 s deposition and 4 s break, of which 4 cycles were done without rotation of the wafer. The QDs formed this way were capped with 2.7 nm GaAs at $\approx$ 500°C. The remaining Indium at the surface was evaporated by ramping up the temperature to 600°C, leaving QDs of 2.7 nm height. 

The QDs were capped with 7.3 nm GaAs and the QD growth step was repeated with the same parameters for the formation of QDMs. This was followed by the tunneling barrier (24.9 nm Al$_{0.34}$Ga$_{0.66}$As) and 166.5 nm of GaAs. Next, the C-doped epitaxial gates were grown, consisting of 10 nm GaAs:C with a doping concentration of $3 \cdot 10^{18}\, / \, \textrm{cm}^{3}$ and 45 nm with $8 \cdot 10^{18}\, / \, \textrm{cm}^{3}$. 
Finally, the sample was capped with 313 nm GaAs for the later structuring of CBGs.

\subsection{Theory}
\label{theory}

Theoretical spectra of positively charged trions in QDM that are shown in Fig.~\ref{fig:Fig_4} are obtained following the approach of Ref.~\cite{Scheibner_2007}. The $X^+$ can be effectively described by a $6\times6$ Hamiltonian
%
%\begin{widetext}
  \begin{align}
  H_{X^+} = E^0_{X^+} \mathbbm{1} +
  \begin{pmatrix}
    -\gamma - 2pF & 0 & t_h & t_h & 0 & 0 \\
    0 & 0 & t_h& t_h & 0 &0 \\
    t_h&t_h& -\Gamma - pF + J^{eh} & 0 & 0 & 0\\
    t_h&t_h& 0 & -\Gamma - pF -J^{eh} & 0 & 0\\
    0& 0& 0& 0& -\Gamma - pF + J^{eh} & 0\\
    0&0& 0& 0& 0& -\Gamma - pF - J^{eh}\\
  \end{pmatrix}
  \end{align}
  which is represented in the basis $\ket{1}=^{\uparrow,0}_{0,\Uparrow\Downarrow}X^+$, $\ket{2}=^{\uparrow,0}_{\Uparrow\Downarrow,0}X^+$, $\ket{3}=^{\uparrow,0}_{\Downarrow,\Uparrow}X^+$, $\ket{4}=^{\uparrow,0}_{\Uparrow,\Downarrow}X^+$, $\ket{5}=^{\uparrow,0}_{\Downarrow,\Downarrow}X^+$, $\ket{6}=^{\uparrow,0}_{\Uparrow,\Uparrow}X^+$ and spin projections are explicitly written for clarity. 
%\end{widetext}
%
Here, $E^0_{X^+}$ is the energy of the trion state with both holes located in the lower QD. $t_h$ denotes the tunnel coupling, $J^{eh}$ the electron-hole exchange interaction strength, $p$ the dipole moment, $F$ the electric field strength and $\Gamma$ ($\gamma$) the energy that is necessary to bring a single (both) hole from the bottom to the upper QD.

Recombination of an electron-hole pair leaves behind a single hole state, which can be represented by two (spin-degenerate) basis states $\ket{1}=^{0,0}_{1,0}h^+$ and $\ket{2}=^{0,0}_{0,1}h^+$:
\begin{align}
  H_{h^+} =  \begin{pmatrix}
    0 & t_h \\
    t_h & -pF
  \end{pmatrix}
\end{align}
Transitions between the six initial trion states and the final hole states that are obtained by diagonalizing $H_{X^+}$ and $H_{h^+}$, respectively, give the PL energies shown in Fig.~\ref{fig:Fig_4} as dashed lines.

\medskip
%\textbf{Supporting Information} \par %Please delete the Suppporting Information statement if it is not applicable. Please supply Supporting Information in another file. Supporting information should not be provided in .tex format
%Supporting Information is available from the Wiley Online Library or from the author.

\medskip
\noindent \textbf{Data availability} \par
\noindent The data that support the findings of this study are available from the corresponding author upon reasonable request.

% Acknowledgements
\medskip
\noindent \textbf{Acknowledgements} \par %delete if not applicable))
\noindent This work was supported by the German Federal Ministry of Education and Research (BMBF) through the Project Q.Link.X. J.J.F. gratefully acknowledges the German Research Foundation (DFG) for financial support via project FI947/6-1. We further acknowledge Technichal Support by the group of Tobias Heindel funded via the BMBF-project ‘QuSecure’ (Grant No. 13N14876) within the funding program Photonic Research Germany.

% References
\medskip

% Use the following code if you wish to generate your bibliography with BibTeX;
% replace the string "MSP-template" below with the name(s) of
% the BibTeX data base(s) you want to use.
% The resulting bibliography-output (the content of the .bbl file)
% must be pasted back into this file before submission.
% Please also include your BibTeX data base file(s) in your submission
% so that we can re-run BibTeX if necessary.
%
\bibliographystyle{MSP}
%\bibliography{MSP-template}

\bibliography{Schall_et_al_QDMs}

\end{document}